\documentclass[prc,floatfix,showpacs]{revtex4}

\usepackage{graphics}
\usepackage{graphicx}

\newcommand{\vs}{\vspace{4mm}}

\newcommand{\be}{\begin{equation}}
\newcommand{\ee}{\end{equation}}
\newcommand{\ba}{\begin{eqnarray}}
\newcommand{\ea}{\end{eqnarray}}
\newcommand{\NL}{\nonumber \\}

\begin{document}


%
%
%

\title{A Non-Extensive Model for Quark Matter Produced in Heavy Ion Collisions}

\author{Tam\'as S. Bir\'o and G\'abor Purcsel}
\affiliation{MTA KFKI Research Institute for Particle and Nuclear Physics,
H-1525 Budapest, P.O.B. 49, Hungary}

\begin{abstract}
We describe quark matter in the framework of non-extensive thermodynamics.
We point out that particle spectra with power-law tail lead to an
increased energy and entropy per particle, and therefore even a massless
plasma may fit both the relatively high value of E/N = 1 GeV and the
spectral slope of T = 175 MeV observed in RHIC experiments.
\end{abstract}


\maketitle


\vs
\section{Introduction}

\vs
The probably most important motivation for relativistic heavy-ion experiments
at large scale accelerator facilities (RHIC, SPS)
is to study quark matter in a unique
environment, possibly in form of a quark-gluon plasma (QGP) produced for a short
while before hadronization \cite{RHICpt}. While the original idea of
quark-gluon plasma sticks to an equilibrium picture both in its early,
bag-model inspired versions \cite{BAGMODEL} and in
the theoretically most advanced lattice gauge models \cite{lattQCD1,lattQCD2}, 
phenomenological
studies of the experimental spectra offer an increasing amount of evidence
that this equilibrium is not completely achieved \cite{CHEMISTRY,ALCOR,PARTONS}.

\vs
In particular the equilibrium thermodynamics relies on (in the Boltzmann
limit) exponential particle spectra, while experiments definitely show
a power-law tail at high transverse momenta. This means that some
assumption leading to the Boltzmann-Gibbs factor is not fulfilled.
We have therefore to look for a theoretical description which is more
general, which releases one or more original assumptions.
Such a candidate is the non-extensive thermodynamics, promoted by
Tsallis and others, which is based on an altered definition of the Boltzmann 
entropy \cite{Tsallis1,Tsallis2,Tsallis3,Plastinos,Tsallis4,Tsallis5}. 
More recently it has been shown that the
power-law distribution also can be described by a logarithmic
dispersion relation for the quasi-particle energy while keeping the
original, extensive entropy definition \cite{QWang1,QWang2}. 
In this case the energy is non-extensive.
The relevance of Tsallis statistics to high-energy $e^+e^-$ and
heavy-ion collisions has been studied in several papers \cite{RELEV}.
In this paper we show that such a version of non-extensive thermodynamics
is able to describe a power-law distribution of quasi-particle energies
and explain an apparent spectral temperature, $T=175$ MeV at low energy. 
Furthermore,
the very power in the tail of this distribution can be connected to
the energy per particle, $E/N=1$ GeV, which remained unexplained in
traditional thermal models.

\vs
There were several attempts to describe dynamical mechanisms which would
simulate (totally or just in part) a Gibbs distribution or a Tsallis
distribution 
with a power-law tail asymptotics 
\cite{BiMu2004,OtherMech1,OtherMech2,OtherMech3,OtherMech4}.
Tsallis advocates the non-extensive thermodynamics based on his
entropy definition for a general use for systems in non-ergodic
states showing exponentially not suppressed distribution tails
\cite{ExplainTs1,ExplainTs2}. Due to this an interesting question arises for the
heavy ion physics: can the quark-gluon plasma (QGP) be described in terms of
the non-extensive thermodynamics? Would it give a quantitatively better
description of experimental observation than the old bag-model picture
while being almost as simple? In this paper we shall answer these
questions positively by showing that a simple QGP with power-law
energy distribution instead of the Gibbs one and a bag constant is able
to reach the hadronically fitted temperature of $T_0 \approx 175$ MeV
and the 1 GeV energy per particle at the same time. We extend this study
for finite chemical potential using relativistic Fermi and Bose
distributions derived according to the rules of non-extensive thermodynamics.
After briefly discussing the recently considered directions in this
research of extending thermodynamics, we choose the interpretation of
Q. Wang \cite{QWang1,QWang2}, who has revealed that an anomalous (logarithmic) 
quasi-particle dispersion relation may lie in the background of the
canonical Tsallis distribution with the famous power-law tail.
We shall study the equation of state (eos) of this QGP
and eventually obtain a constant energy per particle curve on the $T-\mu$ plane.


\vs
\section{QGP in non-extensive thermodynamics}

\vs
The starting point of our description of a quark-gluon matter
at the instant of hadronization is based on a power-law
energy distribution of partons,
\be
 w_1(E) = (1 + E/b)^{-c},
 \label{POWER-LAW}
\ee
containing two parameters; an energy scale $b$ in the order of 1 to a few
GeV and the power $c$, which is expected to be rather high, 4-10, 
extracted from fits to minijet distribution in a pQCD based 
theoretical parton dynamics model \cite{FriesNonaka,Fries}.
The ZEUS collaboration extracted the value $c=5.8 \pm 0.5$
in $e^+e^-$ experiments \cite{ZEUS}.
In the double limit, $c \rightarrow \infty$, $b \rightarrow \infty$
with a fixed ratio $T = b/c$ this distribution approaches the familiar 
canonical Gibbs-Boltzmann formula,
\be
 w_{\infty}(E) = \exp(-E/T).
\ee
As it has been indicated in Ref.\cite{BiMu2004} parton recombination,
which is most likely to happen at low relative energies and therefore
combines a parton cluster or pre-hadron with energy $E$ from two partons with
energy $E/2$, or in general from $n$ partons each with an energy of $E/n$,
drives the power-law distribution towards the exponential limit:
\be
 w_n(E) = (1 + E/(nb))^{-nc},
\ee
by increasing both the effective $c$ and $b$ to $nc$ and $nb$ while keeping
their ratio $T=nb/nc=b/c$ constant. The power-law distribution we are using
here is related to a canonical Tsallis distribution by setting the
temperature $T=b/c$ and the Tsallis index to $q=1+1/c$ when considering
$w^q$-weighted averages of energy and particle number as thermodynamical
observables. Funny enough, using the ''normal'', $w$-weighted averages
as macroscopic variables one arrives at $q=1-1/c$. 

\vs
An anomalous thermodynamics based on an altered definition of the entropy,
\be
 S = \frac{1}{1-q} \sum_i (w_i^q - w_i),
 \label{TSALLIS-ENTROPY}
\ee
has been worked out by Tsallis and several other authors 
\cite{Tsallis1,Tsallis2,Tsallis3,Plastinos,Tsallis4,Tsallis5}.
It resembles
the Legendre transformation structure and herewith the canonical and
grand canonical versions of phase space distributions behind observable
macroscopic expectation values. The distribution (\ref{POWER-LAW}) can be
viewed as the canonical Tsallis distribution
\be
 w_i = \frac{1}{Z} \left( 1 + \frac{(q-1)E_i}{T} \right)^{\frac{1}{1-q}}
 \label{CANONICAL-TSALLIS}
\ee
with the canonical partition function,
\be
 Z = \sum_i \left(1 + \frac{(q-1)E_i}{T} \right)^{\frac{1}{1-q}}.
\ee
The interpretation of the temperature $T$ and the Tsallis-index $q$
(related to the power-low cutoff $b$ and power $c$) is still subject
to the fact, however, that which version of macroscopic observables
are considered in the effective, "non-extensive" thermodynamics.
By now the choice
\be
 E = \sum_i w_i^q E_i / \sum_i w_i^q,
 \label{Q_AVERAGE}
\ee
and the canonical variational principle,
\be
 S - \beta E = {\rm max.}
\ee
seem to establish themselves. (However, a recent article  \cite{QWang2}
righteously criticizes this choice. According to the interpretation given there
the ''normal'' macroscopic averages are meaningful as well.)
In the grand-canonical version $E$ is replaced by $X=E-\mu N$, and $E_i$
by $X_i=E_i-\mu Q_i$. The expression to be minimized becomes $S - \beta X$.


\vs
In the following we review the most important details of these calculations.
Considering the derivatives with respect to $w_i$, 
\ba
 \frac{\partial}{\partial w_i} S &=& c(1-qw_i^{1/c}), \NL
 \frac{\partial}{\partial w_i} X &=& q \frac{w_i^{1/c}}{\sum w_j^q} (X_i-X),
\ea
one arrives at the constraint
\be
 1 = \left( 1 + \beta \frac{X_i-X}{c\sum w_j^q}\right) q w_i^{1/c}
\ee
This gives rise to a grand-canonical distribution $w_i$ proportional
to a $(-c)$-th power:
\be
  w_i \propto \left( 1 + \beta \frac{X_i-X}{c\sum w_j^q} \right)^{-c}.
\ee
This can be casted into the Tsallis form,
\be
  w_i = \frac{1}{Z} \left(1 + \frac{X_i}{cT} \right)^{-c}
\ee
normalized by the partition sum,
\be
  Z = \sum_i \left(1 + \frac{X_i}{cT} \right)^{-c}. 
\ee
The temperature, $T$ is, however, no more related to the Lagrange
multiplier $\beta=1/\Theta$ as straight as usual. It is
\be
  T = \Theta \sum w_j^q  - X/c.
  \label{THETA_T}
\ee
Since the original probability is normalized,
\be
  \sum_i w_i = 1,
\ee
in general $\sum_i w_i^q$ differs from one.


\vs
In order to make the discussion more transparent we use the
following abbreviations for the high-power-law distribution function
and its inverse:
\ba
 \exp_c(x) &= (1-x/c)^{-c} &\longrightarrow  \exp(x), \NL
 \ln_c(x)  &= c(1-x^{-1/c}) &\longrightarrow  \ln(x).
 \label{FUNCTIONS}
\ea
The limits are valid for $c \rightarrow \infty$ by fixed $T=b/c$.
These functions are inverse to each other, $\exp_c(\ln_c(x))=x$, and
reflect the non-extensive property,
\be
 \ln_c(xy) = \ln_c(x) + \ln_c(y) - \frac{1}{c} \ln_c(c)\ln_c(y).
\label{NON-EXTENSIVE}
\ee
The derivatives of these functions follow rules similar to the familiar ones:
\ba
 \frac{d}{dx} \exp_c(x) &=& \exp_c^q(x), \NL
 \frac{d}{dx} \ln_c(x)  &=&  1/x^q. 
\label{DER-RULES}
\ea

The Tsallis entropy with this notation reads as
\be
 S = \sum_i w_i \ln_c(1/w_i),
\ee
and in case of equiprobability, $w_i = 1/N$ for $N$ states, one arrives
at
\be
 S = \ln_c(N), \qquad {\rm and} \qquad N = \exp_c(S).
\ee
The grand-canonical distribution and partition sum becomes
\be
 w _i = \frac{1}{Z} \exp_c(-X_i/T), \qquad
 Z = \sum_i \exp_c(-X_i/T).
 \label{GRAND_TSALLIS}
\ee

\vs
The anomalous thermodynamics reflects familiar looking relations,
with some exceptional points. The heat is expressed by
\be
 \Theta S \sum_i w_i^q \, = \, T \ln_c Z + E - \mu N,
\ee
with the grand canonical partition sum
\be
 Z = \sum_i \exp_c(-(E_i-\mu Q_i)/T),
\ee
where we use the q-weighted particle number,
\be
  N = \sum_i w_i^q Q_i / \sum_i w_i^q,
\ee
and the chemical potential, $\mu$, the corresponding Lagrange multiplier.
Here $Q_i$ are the conserved charges of the state $i$.
The pressure, assuming homogeneity in a sufficiently large volume, $V$,
is given by
\be
 p = \frac{T}{V} \ln_c Z.
\label{PRESSURE}
\ee
Resolving expression (\ref{THETA_T}) for $\Theta$ one arrives at the
following grand canonical potential
\be
 - \Omega = pV = T \ln_c Z = TS-(E-\mu N)(1-S/c).
\ee
The derivatives are given by
\be
 \frac{\partial }{\partial V } \Omega = - p, \qquad
 \frac{\partial }{\partial T } \Omega = - S, \qquad
 \frac{\partial }{\partial \mu } \Omega = - N(1-S/c).
 \label{DERIV_PV}
\ee
For the energy per particle the correction factor $(1-S/c)$, by chance,
does not play a direct role:
\be
  E/N \, = \, \frac{\mu (\partial p/\partial \mu) + T (\partial p/\partial T)-p}{\partial p/\partial \mu}.
\ee

\vs
In case of an ideal (but still power-law, not Gibbs-distributed) gas
one assumes that
\be
 p(\mu,T) = \int \! \frac{d^3k}{(2\pi)^3} \, T \ln_c Z
\ee
with
\be
 Z = \sum_n \exp_c(-nx),
\ee
and
\be
 x_k = E_k - \mu Q_k.
\ee
(Here $Q_k=Q$, the conserved charge is usually independent of the momentum
of the considered particle.) For fermions the summation runs over $n=0$ and
$n=1$ only, while for bosons for all positive integers $n=0,1,2,\ldots$.
In this paper we shall deal with massless relativistic particles
(quarks and gluons), so practically $E_k = |\vec{k}|.$
The derivative relations known for the ordinary ideal gas are almost
fulfilled,
\be
 n = \frac{\partial p}{\partial \mu}, \qquad
 s = \frac{\partial p}{\partial T}, \qquad
 e = \mu n + {\Theta }s - p
\label{IDEAL-DERIV}
\ee
for the particle-, entropy- and energy-density, respectively.

\vs
Constructing this way the Tsallis-Bose and Tsallis-Fermi distributions
(alternative ways can be found in \cite{BEetFDinTS}),
one may use the following integral 
representation\footnote{Alternatively a variable transformation to $x=b/E$ 
projects the interval $(0,\infty)$ to $(0,1)$ and then the trapeze rule or
a higher order Simpson method can be used for numerical integration.
In this case the sensitivity to the finiteness of $dx$ has to be studied
carefully.} of the power-law function:
\be
  \exp_c(-x) = \frac{1}{\Gamma(c)}  \int_0^{\infty} \! dt \,
  t^{c-1} \: e^{-t(1+x/c)}
\ee
where $\Gamma(c)$ is Euler's Gamma function giving $(c-1)!$ for integer
$c > 1$. This representation relates the grand canonical partition function
$Z$ of anomalous thermodynamics to the usual one-particle forms of Fermi
and Bose distributions, respectively:
\be
 Z = \frac{\int_0^{\infty}\!dt\,t^{c-1}e^{-t}\zeta(tx/c) }{\int_0^{\infty} \! dt \, t^{c-1} e^{-t}},
\ee
where $\zeta(y) = 1 + \exp(-y)$ for the Fermi, and $\zeta(y)=1/(1-\exp(-y))$
for the Bose distribution. This form is quite suggestive: the Tsallis
thermodynamics is an average of "normal" systems with an Euler-Gamma distributed
parameter, factorizing the energy. Applying it for $x=E/T$ we substitute
$y=tE/b$, the scaled energy into the "normal" partition function $\zeta$.
This sheds some light on interpretations of Tsallis-distributed particle
spectra in terms of a temperature fluctuating according to the Gamma
distribution \cite{OtherMech3} or energy non-conservation due to higher than
two-body collisions, again with Gamma distributed energy imbalance
\cite{RAFLET,BOLeqinTS}. These correspondences are, of course, no
explanations for the power-law distribution, 
but rather equivalent re-formulations,
assuming the Gamma distribution for the corresponding
(Legendre associated) intensive variable.

\vs
Our main goal in this paper is to investigate the original idea of quark-gluon
plasma in terms of the anomalous, non-extensive thermodynamics. Therefore we
do not wish to speculate here about the microscopic dynamical mechanisms,
which lead to the power-law distribution, closely approximating exponential
spectra in the intermediate transverse momentum range. We assume now, that
an anomalous QGP of massless and Tsallis distributed particles have existed
before hadronization and ask the question whether this state is realistic,
whether it is nearer or farer from the state of hadronic matter observed
in RHIC (and partially in SPS) experiments with many of its properties
surprisingly well fitted
by traditional equilibrium thermodynamical concepts underlying the
hadronic thermal model \cite{ThMOD1,ThMOD2,ThMOD3,ThMOD4}. 
For this purpose we select the
energy per particle $E/N$ as a decisive observable; it can be quite
directly measured on the one hand and quite easily calculated in
thermodynamical framework on the other hand.
Considering a bag constant also in the case of Tsallis QGP, we shall 
modify $E/N$ accordingly. Another interesting quantity is the entropy
per particle, $S/N$. It can be related to other thermodynamical quantities
as
\be
 \frac{S}{N} = \frac{s}{n} = \frac{e+p}{nT} - \frac{\mu}{T}.
\ee

\vs
\section{Energy per particle}

\vs
For calculating $E/N$ it is helpful to introduce notations for the
$(1-S/c)$ uncorrected derivatives of the pressure,
$e_0=T\partial p/\partial T +\mu \partial p/\partial\mu -p$, and
$n_0=\partial p/\partial\mu$. Since the derivatives of the pressure
with respect to $T$ and $\mu$ can be regarded as derivatives of
$\ln_c Z(x)$ with respect to $x$ under the phase space integral by
noting that $\partial/\partial x $ acts as $-T\partial/\partial E_k$
or $T\partial/\partial\mu$ respectively, we get
\ba
 n_0 &=& \int\! \frac{d^3k}{(2\pi)^3} \, f_c({ \frac{E_k-\mu}{T} }), \NL
 e_0 &=& \int\! \frac{d^3k}{(2\pi)^3} \, E_k f_c({ \frac{E_k-\mu}{T} }),
 \label{TSA_INTEGRALS}
\ea
with the one-particle distribution function
\be
 f_c(x) = - \frac{\partial}{\partial x} \ln_c Z(x).
\ee
Now we have simply $E/N = e_0/n_0$.

\vs
Before dealing with Fermi and Bose statistics in the anomalous thermodynamics
let us first review the dilute matter limit, the Tsallis-Boltzmann
approximation. This relies on the assumption that $\exp_c(-x)\ll 1$.
Now $Z=1$ to leading order and there is no difference between Fermi and
Bose distributions,
\be
  f_c(x) \approx - \frac{d}{dx} \ln_c(1+\exp_c(-x)) \approx
  -\frac{d}{dx} \exp_c(-x) = \left( 1 + \frac{x}{c}\right)^{-c-1}.
\ee
A (slight) difference can be observed between the inverse slope
of logarithmic one-particle spectra and the temperature,
\be
  (c+1)T_{{\rm slope}} = cT = b.
\ee
We carry out our first estimate at $\mu=0$ and finite temperature in the
Boltzmann approximation. Using $f_c(E)=(1+E/b)^{-c-1}$ in the corresponding
integrals (\ref{TSA_INTEGRALS}), we get
\ba
  n_0 &=& \frac{d}{2\pi^2} \frac{2b^3}{c(c-1)(c-2)},  \NL
  e_0 &=& \frac{d}{2\pi^2} \frac{6b^4}{c(c-1)(c-2)(c-3)}, 
\ea
with a common (color, spin, etc.) degeneracy factor $d$. In the low energy
region we obtain a simple relation between the energy per particle and the
inverse slope of the one-particle spectra,
\be
  E/N = e_0/n_0 = \frac{c}{c-3} 3T = \frac{3(c+1)}{(c-3)} T_{{\rm slope}}.
  \label{E_PER_N_BOLTZMANN}
\ee
This differs from the familiar relation, $E/N = 3T$, known for the 
massless Boltzmann gas. It is interesting to note, that -- as it
has been calculated in Ref.\cite{Lavagno1,Lavagno2} 
-- by using the normal weighted
definition for particle number, $N'=\sum_i w_iQ_i$, one gets
$E/N'=3T$\footnote{For normal weighted quantities 
$f_c(x)\approx (1+E/b)^{-c}$, therefore $n_0'=(d/2\pi^2)2b^3/(c-1)(c-2)(c-3)$.}. 
We do not think, however, that using different definitions
for the macroscopic energy and particle number would make a sense
in comparison with experimental data: either $w_i$ or $w_i^q$ is
the distribution of physical states. Furthermore $T$ is not exactly
the "experimental temperature" conjectured from the inverse spectral slope.

\vs
An estimate of the inverse slope gives rise to $T_{{\rm slope}} \approx 175$
MeV for RHIC experiments in the $\mu \approx 0$ region. This is also the
value used by conventional thermal models for the temperature.
Aiming at $E/N = 1$ GeV now, the formula (\ref{E_PER_N_BOLTZMANN}) 
leads to a power of $c+1 \approx 8.42$ in the one-particle
distribution $f_c(E)$, a value
comparable with minijet $p_T$-distribution fits \cite{Fries,FriesNonaka}.
Neither the traditional Boltzmann formula, $E/N=3T$, nor the simple bag model
of QGP, where $E/N=4T$, 
is able to describe both this inverse slope and energy per particle
at the same time. 

\vs
In order to display a more throughout comparison to the thermal model
points in the  $T-\mu$ plane fitted to different accelerator experiments
\cite{StachREVIEWS}, we carried out numerical
integrations with Tsallis-Fermi and Tsallis-Bose distributed, massless
quark-gluon plasma at finite temperature and chemical potential.
In this case, denoting a general integral of $f_c(x)$ weighted with
a power $x^n$ by
\be
 I_n(a)= \int_{-a}^{\infty} dx \, x^n \, f_c(x),
\ee
and the symmetric and anti-symmetric combinations, $S_n(a)=I_n(a)+I_n(-a),$
$A_n(a)=I_n(a)-I_n(-a)$, we arrive at
\ba
 2\pi^2 \: n_0 &=& T^3 S_2(\mu/T) + 2T^2 \mu A_1(\mu/T) + T\mu^2 S_0(\mu/T), \NL
 2\pi^2 \: e_0 &=& T^4 S_3(\mu/T) + 3T^3 \mu A_2(\mu/T) + 3T^2\mu^2 S_1(\mu/T)
 		+ T\mu^3 A_0(\mu/T), \NL
 & &	\quad	
\ea
for each fermion. For the boson part we assume
$\mu=0$. The energy per particle characteristic for the power-law
distributed mixture then becomes
\be
  E/N = \frac{d_Be_{0B}(T)+d_Fe_{0F}(T,\mu)+d_Fe_{0AF}(T,\mu)}{d_Bn_{0B}(T)+d_Fn_{0F}(T,\mu)+d_Fn_{0AF}(T,\mu)}.
\label{EQ43}
\ee
Before presenting our numerical results there is one more aspect of this version
of the anomalous thermodynamics to be discussed. In the traditional, Gibbs
thermodynamics the anti-fermion distribution is related to the
fermion distribution simply by a change in the sign of the baryochemical
potential, $\mu$. This formula exactly coincides with the density of
fermion holes:
\be
  1 - f_{\infty}((-E-\mu)/T) = f_{\infty}((E+\mu)/T)
\ee
due to $f_{\infty}(x) + f_{\infty}(-x) = 1$ for the Fermi distribution
$f_{\infty}(x) = 1/(1+\exp(x))$. This relation is not fulfilled for
a finite $c$ index Tsallis-Fermi distribution: 
\be
 f_c(x) = (1+\exp_{-c}(x))^{-q}.
\ee
In this case
\be
 f_c((-E-\mu)/T) + f_c((E+\mu)/T) = f_c(-x)+ f_c(x) \ne 1,
\ee
instead
\be
 f_c^{1/q}(x) + f_{-c}^{1/q}(-x) = 1
\ee
is fulfilled.
Here the anti-charge choice with $-\mu$ and the particle hole choice
by replacing $-E$ energy are not equivalent.
For the estimate of the energy per particle of a Tsallis QGP at
finite temperature $T$ and baryochemical potential $\mu$ in 
Ref.\cite{RHIC-SCHOOL}
we utilized the particle-hole correspondence:
\be
\overline{f}_c(E;\mu,T) = 1 - f_c(-E;\mu,T).
\ee
Now we turn to another interpretation of the non-extensive
thermodynamics, similar to that suggested by 
Q. Wang \cite{QWang1,QWang2,EXTversionTS}, 
which handles the fermions and antifermions in a more satisfactory manner.

\vs
\section{Quasiparticles and anomalous thermodynamics}

\vs
In the preceding discussion we have already indicated 
some disadvantages of the ''main stream''
Tsallis formulation of non-extensive thermodynamics. 
These make it to be not the best choice for our purpose, i.e.
for using it in the description of quark matter. The dependence of
some thermodynamical relations indirectly on the system size
(i.e. dependence of densities on the total entropy, etc.), the slight difference
between temperature parameter and one-particle distribution inverse slope,
and finally the strange correspondence between particles and holes
in the Fermi distribution leave us with dissatisfaction. 
Also the very basic point, the non-extensivity
of the Tsallis-entropy remains a challenge for the physical interpretation
(and acceptance) of this approach.

\vs
Fortunately it has been shown meanwhile that a mathematically well-defined
monotonous function of the Tsallis entropy describes an extensive entropy
measure \cite{Abe,QWang1}. The Tsallis-entropy satisfies the general
rule of pseudo-additivity,
\be
 f(S_{12}) = f(S_1) + f(S_2) + \lambda f(S_1) f(S_2),
\ee
required by the existence of thermal equilibrium in non-extensive composite
systems. The Tsallis-entropy realizes the case $f(S)=S$ and $\lambda=1/c=q-1$.
The additive, i.e. extensive measure of entropy then becomes
\be
\tilde{S} \, = \, \frac{1}{\lambda} \: \ln (1 + \lambda f(S) ).
\ee
Based on the maximalization of $\tilde{S}$ of the Tsallis-entropy instead
of $S$ itself, a formally traditional thermodynamics and canonical, or
grand canonical distribution emerges \cite{QWang2}. The price to pay for
that is, that the energy (and incidentally the particle number and the volume)
is no more an extensive quantity. It is a clear indication that the Tsallis
entropy describes systems whose parts still interact with long range
forces. The Tsallis canonical distribution and partition sum can be
transformed into an ordinary canonical partition sum using a
''deformed Hamiltonian''. By defining
\be
\tilde{H} = \frac{1}{\beta (q-1)} \ln \left(1 + (1-q)\beta(H-\mu N) \right)
\ee
one arrives at
\be
 Z = {\rm Tr} \left(1-(q-1)\beta(H-\mu N) \right)^{1/(q-1)} 
   = {\rm Tr} e^{-\beta \tilde{H}}.
\ee
As a consequence the Fermi and Bose distributions keep their traditional
forms as functions of the eigenvalues of the deformed Hamiltonian, $\tilde{H}$.
The trace over the states of the many particle system leads to a product
over the phase space labels of independent quasi-particles with energy
$\omega_k$, and the one-particle distribution becomes
\be
 f(k) = \frac{1}{e^{\omega_k/T}\pm 1} = 
  \frac{1}{(1+(E_k-\mu)/cT)^c \pm 1}.
\ee
where $E_k$ is the eigenvalue of the original Hamiltonian, i.e.
the energy of an interacting quark or gluon with momentum $k$.
For particle and nuclear physics purposes this use of a quasi-particle
energy is the more enlightening approach.

\vs
For the investigation of a long range interacting quark matter
we still have to modify the above picture a little. The power-law
tail with a general, fractional power restricts the base of this
power-law to be positive. In a picture of a QGP
rooting in QCD on the other hand  gluons are exactly and quarks are
practically massless, therefore $E_k=|k|$ can be as small as zero.
For any finite, positive chemical potential $\mu>0$ the above distribution
may become complex, unless $\mu < cT = b$. This restricts the use of
quark chemical potentials too seriously. We explore therefore here another
approach.

\begin{figure}
\begin{center}
 \includegraphics[width=0.45\textwidth]{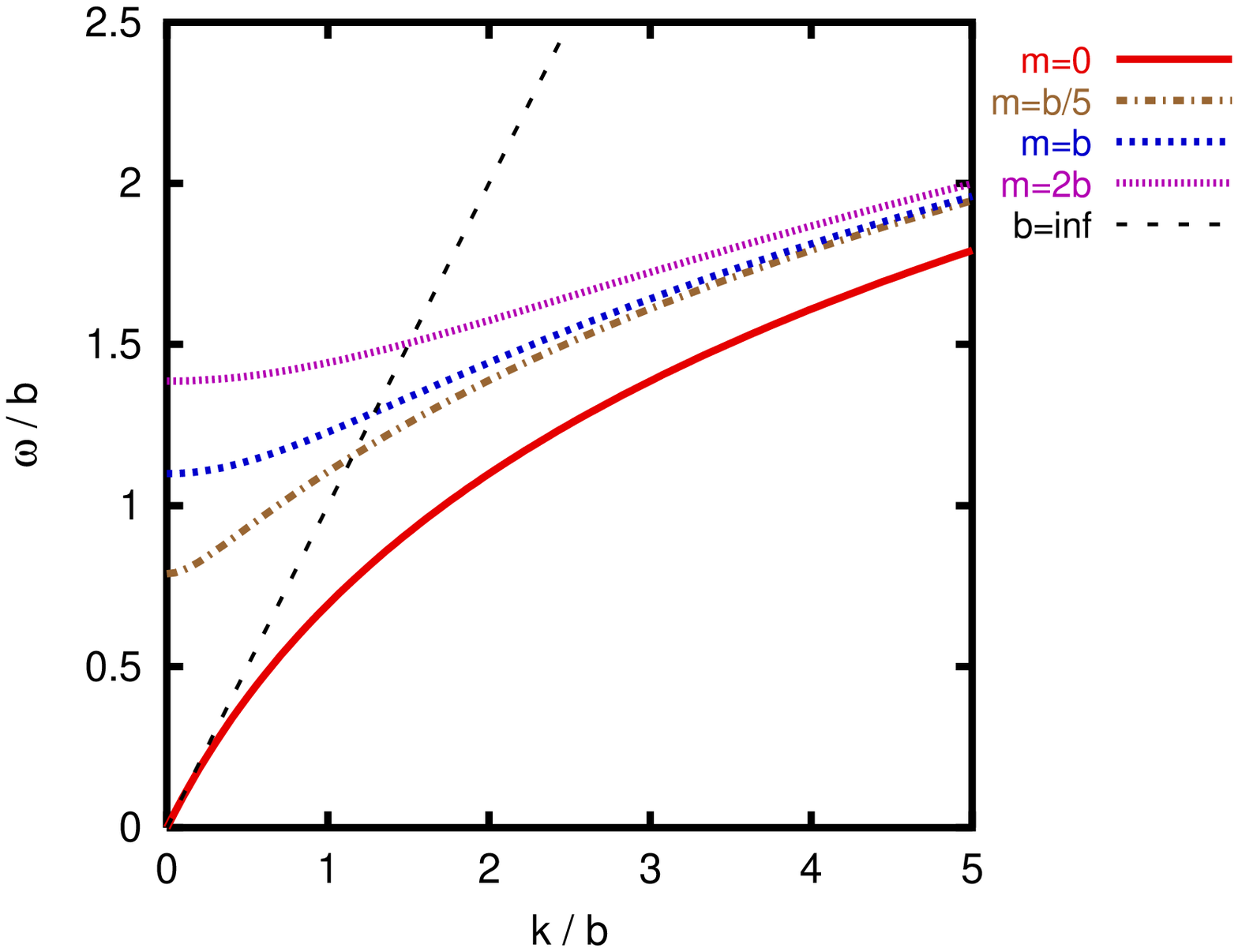}\includegraphics[width=0.45\textwidth]{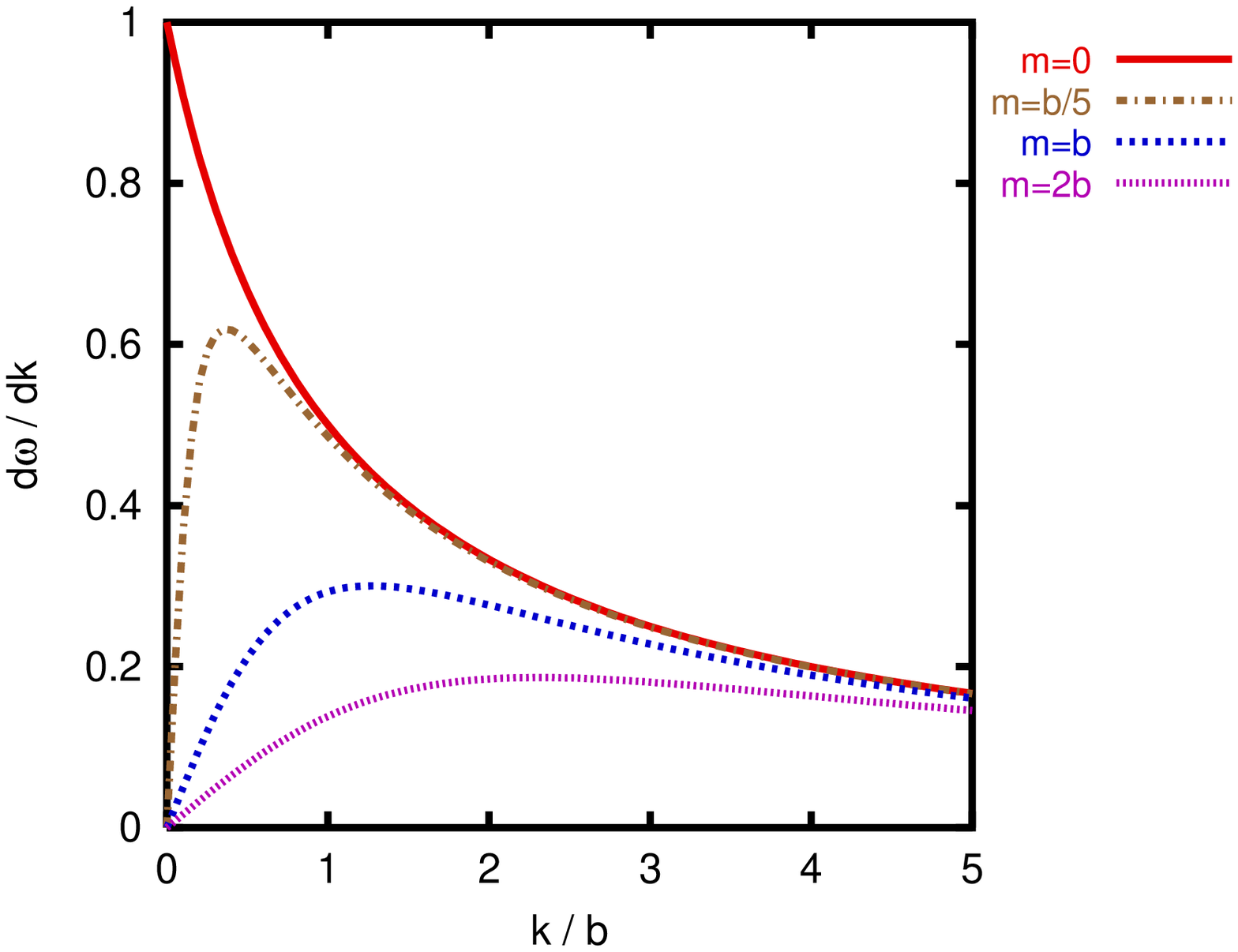}
\end{center}
\caption{Anomalous dispersion relations and group velocities for Tsallis quark matter.}
\label{DISPREL}
\end{figure}


\vs
Repeating the above line of derivation given by Q.Wang we start with
the anomalous dispersion relation for the QGP quasi-particles:
\be
 \omega_k = b \ln (1 + E_k/b )
 \label{BIRO}
\ee
with $E_k=\sqrt{k^2+m^2}$ being the dispersion relation of a free
relativistic particle with mass $m$. Fig.\ref{DISPREL} shows this
for different masses. The right side of the figure plots the
group velocity, $v_g = d\omega/dk$, which overall remains
below one. Such a dispersion relation does not violate causality.
Still it is anomalous; it shows an acoustical branch 
\cite{LANDAU,SOLIDtxtbook} for the massless
case, and a line crossing from the optical branch at low momenta to
an acoustical behavior at high momenta for finite mass. Such a 
quasi-particle dispersion relation has never been considered yet in QCD,
as far as we know \cite{QUASI1,QUASI2,QUASI3}. 
The plasmons all live on the optical branch, above
the light cone. Below the light cone a series of Landau poles can be
located, they cannot belong to traditional ''particles''. It is also
hard to find a (partial) summation of diagrams leading to a logarithmic
dispersion relation from QCD. But the proof of impossibility is also
unknown\footnote{For very high momenta this dispersion relation also
becomes invalid: one expects a restoration of the isotropy between
spacelike and timelike four-momentum components.}.

\vs
The investigation of the non-relativistic and extreme relativistic limit
of the formula (\ref{BIRO}) may help to approach to a physical interpretation.
Based on the second derivative,
\be
  \frac{d^2\omega_k}{dk^2} \, = \,
  \frac{m^2/\omega_k +(m^2-k^2)/b}{(\omega_k+\omega_k^2/b)^2}
\ee
we get in the non-relativistic range $0 \ll k \ll m $ 
(which nevertheless cannot be applied to massless QGP) 
\be
 \omega_k = b \log(1+\frac{m}{b}) + \frac{b}{b+m}\frac{k^2}{2m} + \ldots
\ee
This features an attractive mean field effect for the heavy quarks in QGP
due to $\omega_0 < m$, and an increased effective mass $ m_{eff} = m(1+m/b)$.
For relativistic (including $m=0$) particles the quasi-particle
mass cannot be interpreted other than the inverse of the second derivative
of the dispersion relation at $k=0$, which is zero as defined
above. At any finite $k$ the second derivative is negative, not
corresponding to any picture of a traditional point particle.

\vs
It is also interesting to interpret this dispersion relation in the
framework of plasma physics. In plasmas a dielectric coefficient,
$\varepsilon(\omega,k)$, may alter the free dispersion relation. The pole
of the propagator in medium is located at
\be
 D^{-1} = \omega^2 - k^2/\varepsilon = 0.
\ee
From this we get
\be
 \sqrt{\varepsilon} = \frac{k}{\omega_k} = \frac{k/b}{\ln(1+k/b)}
 \label{EPSILON}
\ee
which is greater or equal to one. This is a further anomalous property,
related to a negative self-energy 
(due to $D^{-1}=\omega^2-k^2-\Pi$, $\Pi=k^2(1/\varepsilon-1)$ ). 
This property also occurs in non-abelian gauge theories, where
gluon loop contributions render the gluon self energy negative.
This effect is alike asymptotic freedom, a vanishing charge at
high momenta. Classically $\alpha = \alpha_0/\varepsilon$, according to
eq.(\ref{EPSILON}) this quantity tends to zero at $k \rightarrow \infty$.
The very functional form of this running coupling, however, differs from the
usual 1/log one, known from perturbative QCD.

\begin{figure}
\begin{center}
 \includegraphics[width=0.4\textwidth]{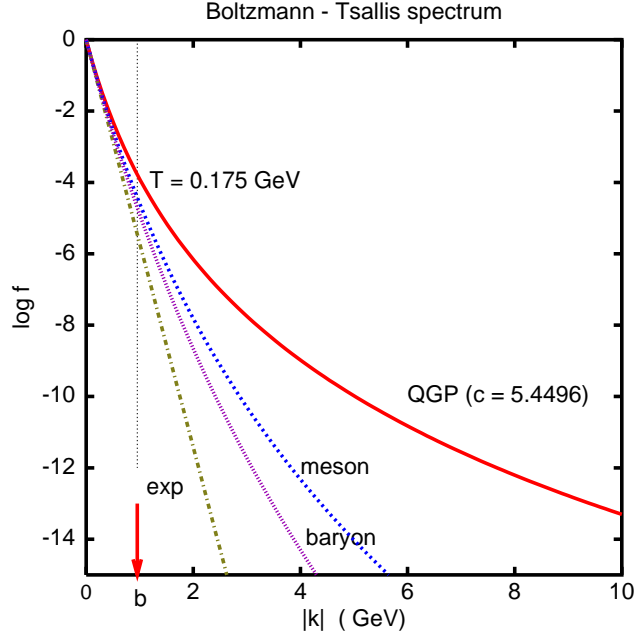}
\end{center}
\caption{One-particle distributions in Tsallis-QGP. Quarks or gluons
correspond to $c=5.4496$, mesons to $2c$ and baryons to $3c$ according
to the recombination idea. An exponential fit to $k=0$ is also indicated.}
\label{TSPEC}
\end{figure}

\vs
The one-particle distribution functions (Fermi and Bose) in the
Boltzmann approximation become simple power-laws:
\be
 f(k) = \exp(-(\omega_k-\mu)/T) =
 \exp(\mu/T) \left(1+\frac{E_k}{b} \right)^{-b/T}
\ee
with a power $c=b/T$. Now the chemical potential $\mu$ does not enter
under a logarithm, which makes it possible to obtain real results at
any value of it. The Fermi and Bose distribution becomes
\be
 f(k) = \frac{1}{\exp{(\omega_k-\mu)/T}\pm 1}
      = \frac{1}{\exp(-\mu/T) \left( 1 + E_k/b \right)^c \pm 1}.
\ee
In the $c \rightarrow \infty$ limit the equilibrium thermodynamics of an
ideal gas is re-established. Otherwise a power-law tail occurs for momenta
around and higher than $b$. Fig.\ref{TSPEC} shows these distributions
for the value $b=1.0325$ GeV, favored by the RHIC experiment slope
$T=175$ MeV and an energy per particle of $E/N = 1$ GeV.

\vs
This anomalous behavior can be traced back to one feature: the
familiar exponential equilibrium distribution has been modified
at high momenta. This contradicts to the traditional result from
perturbative QCD, so it cannot stem from an interaction
correction to the static, equilibrium state at finite temperature.
We rather think it may be related to a characteristically short time
scale preventing equilibrium. The short-time behavior is somehow
non-particle like, probably related to high off-shell-ness of
emerging hadrons.

\vs
Recovering the power-law in the Tsallis-Boltzmann approximation, the
important thermodynamical quantities coincide with our previous results [ref RHIC].
An other important limit, the $T=0$ Fermi distribution also can be
calculated analytically for massless quark matter using the logarithmic
dispersion relation (\ref{BIRO}). The contributions to pressure,
energy density and particle density follow the traditional
relation, $e=Ts+\mu n-p.$ Considering a bag constant, $e=e_0+B$ and
$p=p_0-B$ still satisfy this relation. The stability edge line, where $p=0$,
can be realized by $p_0=B$, and therefore $e=e_0+p_0$. 
On this edge of the QGP stability against clustering, -- possibly a state
at the hadronization process, -- the energy per particle and the entropy
per particle are linearly related for any ideal quasi-particle system.
In particular at $T=0$, $E/N = \mu$ has to be fulfilled.
As a consequence a hadronic thermal model with $E/N = 1$ GeV would
have to end at $\mu_B = 1$ GeV. This estimate
does not leave much room for $\mu = \mu_B/3$ massless quarks as immediate
precursors of the hadrons at the same energy per particle at $T=0$.

\vs
Direct integration under the Fermi energy $\mu$ gives the following
results at zero temperature:
\ba
 p_0 &=& \langle \frac{k}{3} \frac{d\omega}{dk} \rangle =  
 \frac{b^4}{6\pi^2} \left(x-\frac{x^2}{2}+\frac{x^3}{3} - \ln(1+x) \right) \NL
 e_0 &=& \langle \omega \rangle =  \mu n - p_0 \NL
 n   &=& \langle 1 \rangle = \frac{b^3}{2\pi^2} \frac{x^3}{3}  .
\ea
with $x = k_F/b$.
The Fermi momentum can be obtained by inverting the logarithmic
dispersion relation, $k_F=b(e^{\mu/b}-1)$. Considering baryons
made of three quarks, finally we set $\mu_B=3\mu$. In the $b \rightarrow \infty$
limit, realizing that the pressure subtracts from the first three 
terms in the Taylor expansion of $\ln(1+x)$ exactly $\ln(1+x)$, 
it is easy to see that $p_0=\mu^4/(24\pi^2)$.
In this small $x$ ($k_F \ll b$, dilute, cold quark matter) limit
$e_0/n \approx b(3x/4 + 9x^2/10 + x^3/6 + \ldots)$, so the pure
energy per particle is $e_0/n = 3\mu/4$.

\vs
\vs
\section{Numerical results}
\vs

\begin{figure}
\begin{center}
 \includegraphics[width=0.35\textwidth]{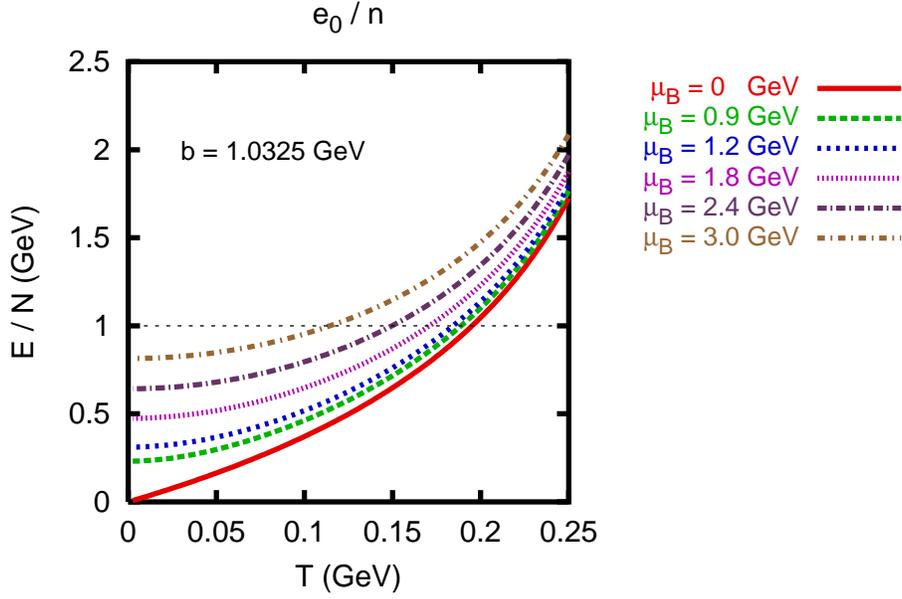}
\end{center}
\caption{Energy per particle in Tsallis-QGP without a bag constant.}
\label{ENOBAG}
\end{figure}

Finally we present results of numerical integrations at arbitrary
$T$ and $\mu$ with logarithmic dispersion relation for massless
quark matter. First, Fig.\ref{ENOBAG} plots the direct energy per particle
$e_0/n$ as a function of temperature for different chemical potentials,
utilizing the $b=1.0325$ GeV value. The 
requirement of having $E/N=1$ GeV at $\mu=0$ in this case leads 
to $T=195$ MeV. Alternatively $T=175$ MeV is achieved by the choice
of $b=0.8275$ GeV. 

\begin{figure}
\begin{center}
 \includegraphics[width=0.35\textwidth]{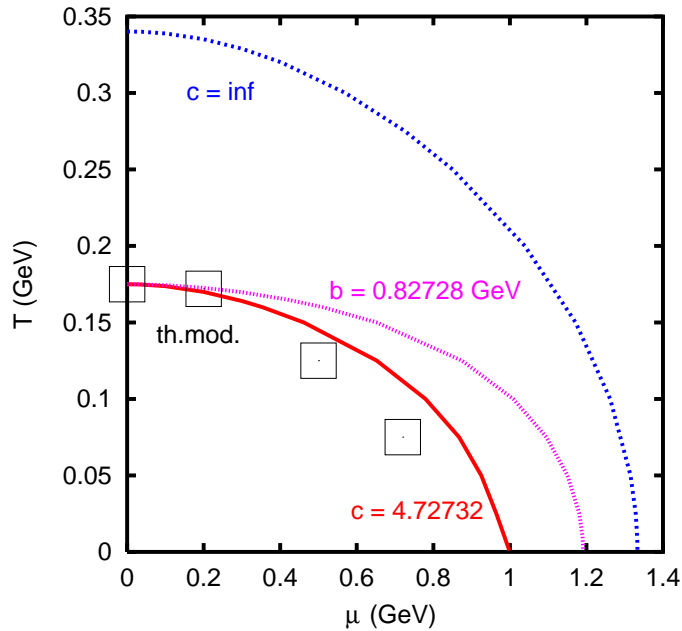}
\end{center}
\caption{The 1 GeV energy per particle line without bag constant in the
$T-\mu$ plane. The constant power, $c=4.72732$, line is solid, the
constant energy parameter, $b=0.82728$ GeV line is dashed-dotted, the
traditional thermal line ($c=\infty$) is dotted. Boxes show corresponding
values from the hadronic thermal model.}
\label{DIRECTLINE}
\end{figure}

\vs
The direct $e_0/n=1$ GeV lines on the $T-\mu$ plane are shown in
Fig.\ref{DIRECTLINE}.The constant power, $c=4.72732$ line (solid line) 
approaches the data of the hadronic thermal model (boxes) most
closely, however only with the $\mu=\mu_B$ assumption. This is not
readily a quark matter, where $\mu=\mu_B/3$ would be. Massless
baryon clusters as prehadrons satisfy rather this assumption, but
this idea -- however exotic on its own right -- does not have yet
any support from microscopic models.
The constant dispersion relation with $b=0.82728$ GeV plots another
line. This deviates even more at $T=0$ from the hadronic boxes,
contrary to the fit to the same point at $\mu=0$. (Both the $c$ and
the $b$ value were obtained from $E/N=1$ GeV at $\mu=0$ and $T=0.175$ GeV.)
Finally the traditional, massless ideal gas corresponding to
the exponential spectra ($c=\infty$) are vastly off from the experimental
data.

\vs
The entropy per particle is a sensitive observable,
because cannot decrease in spontaneous hadronization processes unless
there are more hadrons than quarks and gluons before.
It is plotted in the massless ideal Tsallis-gas scenario along
the $E/N = 1$ GeV curve as a function of the temperature.
The solid line corresponds to a constant $c=4.72732$. It is worth
to note that it ends at a finite value at $T=0$, contradicting to the
third law of thermodynamics. Probably the $c=$constant assumption
is not realistic. (Of course this deviation is proportional to $1/c$
as it is easy to see from a low-$T$ expansion of the Fermi distribution.)
The constant $b=0.82728$ GeV curve (dashed-dotted) and the traditional
thermodynamics (dashed) curve both start from zero at $T=0$.
However, only the finite $b$ case can fit the high value
of $S/N \approx 6.7$ at $T=0.175$ GeV dictated by the $E/N = 1$ GeV
for massless bosons and fermions.

\begin{figure}
\begin{center}
 \includegraphics[width=0.35\textwidth]{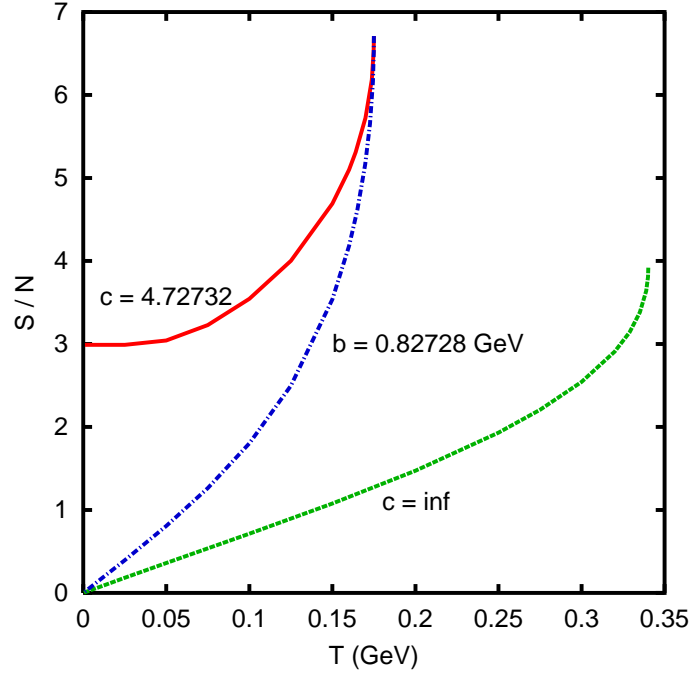}
\end{center}
\caption{Along the 1 GeV energy per particle line without bag constant in the
$T-\mu$ plane the entropy per particle, $S/N$, is plotted. 
The constant power, $c=4.72732$, line is solid, the
constant energy parameter, $b=0.82728$ GeV line is dashed-dotted, the
traditional thermal line ($c=\infty$) is dashed. 
}
\label{SNDIRECT}
\end{figure}

\begin{figure}
\begin{center}
 \includegraphics[width=0.3\textwidth]{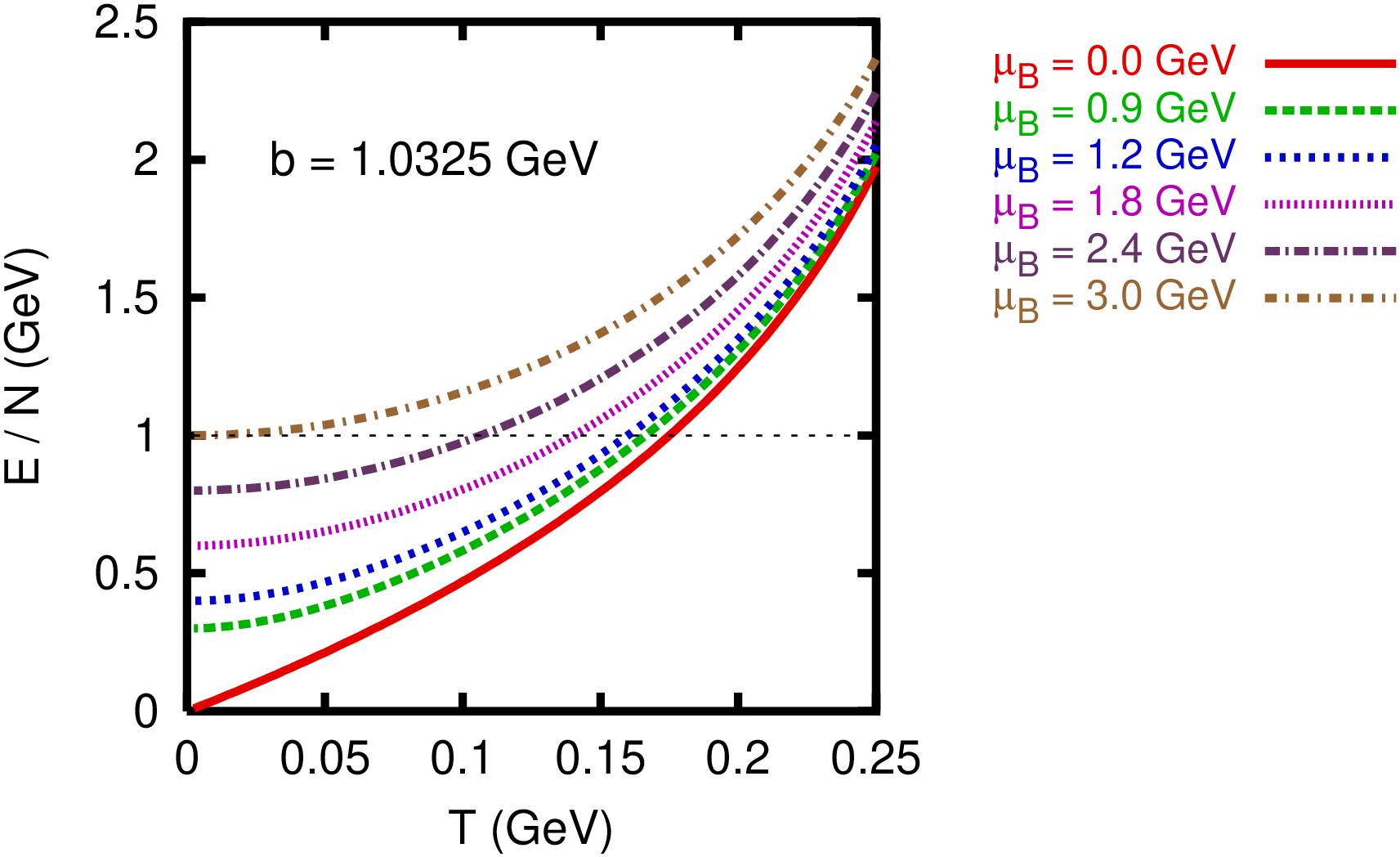}\includegraphics[width=0.3\textwidth]{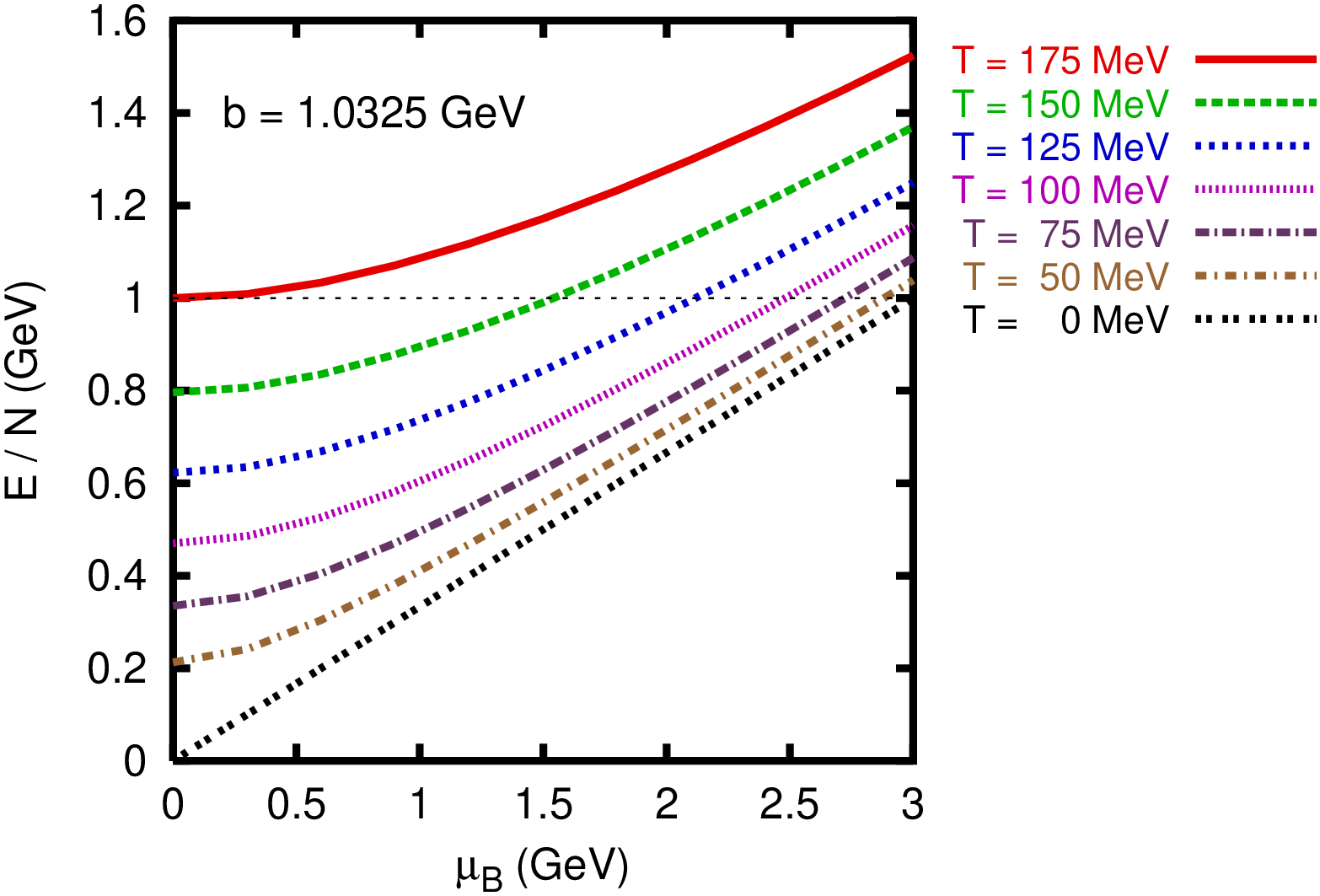}
\end{center}
\caption{Energy per particle in a Tsallis-QGP with adjusted bag constant
 to the $p=0$ line. Isochors (left) and isotherms (right) are shown.}
\label{ENvsT}
\end{figure}

\vs
Taking into account a possible bag constant in Tsallis-QGP
with logarithmic ''quasi-particle'' energy, the fitted $b$ value
becomes $1.0325$ GeV. In Fig.\ref{ENvsT} the (kinetic + bag) 
energy per particle, $(e_0+p_0)/n$ at the $p=0$ stability limit
is plotted as function of the temperature, $T$ (left part), and 
as a function of the chemical potential, 
$\mu_B=3\mu$ (right part).

\vs
Finally we present the $p=0$ stability line for a constant $b=1.0325$ GeV
dispersion relation, while adjusting the bag constant $T$ and $\mu$
dependent so, that the total energy per particle, $E/N=(e_0+p_0)/n$
remains constant, $1$ GeV. We find that of all Tsallis-distributed
massless QGP models this prescription may come
the closest to the hadronic thermal model data. 
Still the low $\mu_B$ value of the hadronic fits to exponential
spectra cannot be achieved,
pointing out that massless QGP at lower-energy heavy-ion experiments is
probably not the state of matter at hadronization, not even with
power-law high-momentum tail. 
Instead massive quarks and gluons have to be considered, which would allow
to reach high $E/N$ and $S/N$ values with high-$p_t$ powers closer
to the experimental values.  At RHIC energies, however, this picture
is promisingly consistent.

\begin{figure}
\begin{center}
 \includegraphics[width=0.35\textwidth]{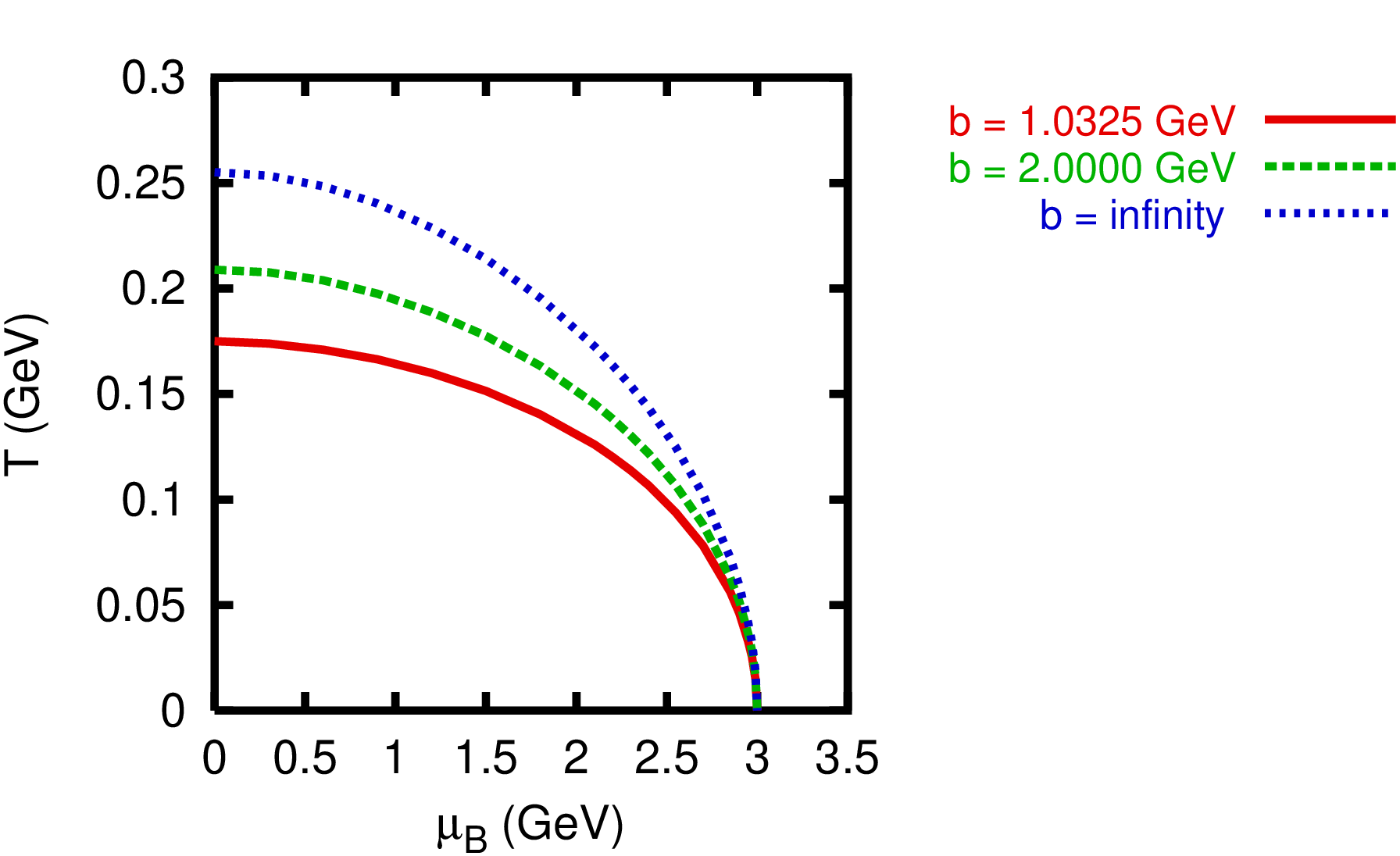}
\end{center}
\caption{The $E/N=1$ GeV line with $p=0$ adjusted bag constant on the
$T-\mu_B$ plane ($\mu_B=3\mu$ assumed).}
\label{BAGLINE}
\end{figure}



\vs
\section{Conclusion}

\vs
In conclusion we have worked out the thermal properties of a plasma of massless
quarks and gluons utilizing a power-law energy distribution and the
Tsallis type anomalous thermodynamic rules. This model describes a
non-equilibrium situation or a non-ergodic stationary state: 
some correlations are not suppressed exponentially, one particle distributions
show a long tail in energy and hence in transverse momenta.

\vs
This is exactly the case which has been observed in relativistic heavy ion
collisions at RHIC at high transverse momenta. The Tsallis QGP is able
to offer a temperature and energy per particle value which are near to
the ones fitted by hadronic thermal models; a property not shared by
traditional quark gluon plasma models. This way a natural precursor
state may be suspected at hadronization, as well as an explanation may be
given for the value $E/N \approx 1$ GeV. Furthermore we have extended this
estimate for finite baryochemical potential using Fermi-Tsallis and
Bose-Tsallis statistics for quarks and gluons respectively. Also in this case
one can come near to the hadronic thermal model curve with certain power
fits, however, due to some indefiniteness in the treatment of anti-fermions
the results are less convincing.

\vs
A possible solution to this problem is offered by introducing a monotonous
re-scaling of Tsallis entropy, which leads to a straightforward
correspondence between traditional and Tsallis distributions.
The key to this is a logarithmic dispersion relation for the quasi-particles,
which has an anomalous character, as we have discussed. This leads to
an anomalous thermodynamics, in which now the anti-particles can be
interpreted as holes in the particle distribution also for the extended,
non-extensive Tsallis distribution. We have repeated the numerical
analysis of the energy per particle and the spectra for this case, and
realized that the basic qualitative features do not change.

\vs
Our final conclusion from these studies is that in order to interpret
experimental observations on the bulk particle spectra in relativistic
heavy-ion collisions we need to use at last generalized thermodynamical
concepts instead of the traditional ones. In particular only this approach
allows an interpretation of a spectral slope, a high tail with power-law
and the fitted $E/N \approx 1$ GeV energy per particle on the QGP side
at the same time.

\vs
It remains to clarify what mechanism leads to an anomalous statistics
in these heavy ion collisions and to explain the very power. This is a
task on its own right, here we just try to sketch some ideas about
a possible explanation. Considering non-ergodic, but stationary solution
of a Boltzmann equation for partons, special energy dependent cross sections 
may lead to an approximate validity of simple recombination rules 
\cite{ALCOR,BialCOALESC,FriesNonaka}.  The power-law distribution recombines in this case
to distributions closer and closer to the familiar exponential \cite{BiMu2004}.
Analog to the ''canonical suppression'', which considers effects due to a finite
volume on the grand canonical distribution \cite{CANSUP,StachREVIEWS}, 
here we aim to consider
finite-time effects leading to partially-equilibrated distributions.

\vs
{\bf Acknowledgment} This work has been supported by the
Hungarian National Research Fund, OTKA (T034269). Enlightening discussions
with J. Zim\'anyi are gratefully acknowledged.


\end{document}